\documentclass[aps,prl,reprint]{revtex4-1}
\usepackage{times}
\usepackage[colorlinks=true, urlcolor=blue, linkcolor=red, citecolor=blue, pdftex]{hyperref}
\usepackage{graphicx}
\usepackage{amsmath}
\usepackage{amssymb}
\usepackage{color}

\def\ket#1{\left|#1 \right\rangle}

\def\braket#1#2{\left\langle #1 | #2 \right\rangle}
\def\matrix22#1#2#3#4{\left(\begin{array}{cc}#1&#2\\#3&#4\end{array}\right)}

\begin{document}
\title{Nature of chiral spin liquids on the kagome lattice}
\author{Alexander Wietek}
\email{alexander.wietek@uibk.ac.at}
\author{Antoine Sterdyniak}
\author{Andreas M. L\"auchli}
\affiliation{Institut f\"ur Theoretische Physik, Universit\"at Innsbruck, A-6020 Innsbruck, Austria}

\begin{abstract}
We investigate the stability and the nature of the chiral spin liquids which were recently uncovered in extended Heisenberg models on the kagome lattice. Using a 
Gutzwiller projected wave function approach -- i.e.~a parton construction -- we obtain large overlaps with ground states of these extended Heisenberg models. We further
suggest that the appearance of the chiral spin liquid in the time-reversal invariant case is linked to a classical transition line between two magnetically ordered phases.
\end{abstract}

\maketitle
\paragraph{Introduction ---}
The quest for quantum spin liquids~\cite{balents2010spin} is currently a very active endeavour in condensed matter physics. This elusive state
of quantum matter comes in various forms and is theoretically intensely studied, however was difficult to pin down in computational studies of {\em realistic} 
quantum spin Hamiltonians and hard to characterise unambigously in experiments on quantum magnets.

The $S=1/2$ Heisenberg antiferromagnet on the kagome lattice has emerged as one of the paradigmatic systems where quantum spin liquid phases are
expected. A plethora of theoretical proposals have been put forward, ranging from valence bond crystals~\cite{Marston1991,Nikolic2003,Singh2007,Poilblanc2011,Capponi2013b}, algebraic spin liquids~\cite{Hastings2000,Ran2007,Hermele2008,Iqbal2013}, 
$\mathbb{Z}_2$ spin liquids~\cite{Sachdev1992,Moessner2001,Balents2002,Misguich2002,Wang2006,Lu2011,Iqbal2011}, to chiral spin liquids~\cite{Marston1991,Messio2012b,Messio2013}. Despite tremendous theoretical and computational progress~\cite{Leung1993,Lecheminant1997,Waldtmann1998,Capponi2004,Jiang2008,Schwandt2010,Yan2011,Laeuchli2011,Jiang2012,Depenbrock2012,Nishimoto2013,Clark2013,Capponi2013}, the true nature of the ground state and the low-lying excited states of the nearest neighbour Heisenberg model on the kagome lattice is still not settled completely.

Chiral spin liquids (CSL) are a particular family of spin liquids in which time-reversal symmetry (TRS) and parity symmetry are (spontaneously or explicitly) broken~\cite{Wen1989a,Wen1989b}. The scalar chirality  $\langle\vec{S}_i\cdot(\vec{S}_j\times \vec{S}_k) \rangle$ is non-zero and uniform and manifests the breaking of time-reversal and parity symmetries, analogous to the presence of an orbital magnetic field. In a favorable situation the breaking of these symmetries could conceivably lead to a spin analogue of the Fractional Quantum Hall Effect, although other types of ground states are possible as well~\cite{Momoi1997,Lauchli2003}. Historically Kalmeyer and Laughlin envisioned such a scenario by considering lattice versions of the bosonic $\nu=1/2$ Laughlin wave function as candidate ground state wave functions for the triangular lattice Heisenberg model~\cite{Kalmeyer1987,Kalmeyer1989}. 

In two recent papers~\cite{Gong2014,Bauer2014}, two forms of chiral spin liquids have been discovered, which are stabilised away from the nearest neighbour
Heisenberg model upon adding further neighbour Heisenberg interactions or scalar chirality terms to the Hamiltonian. Both studies numerically demonstrate the required ground state degeneracy and characterize the underlying topological order by computing the modular matrices.

This breakthrough lays the foundation for further investigations of chiral spin liquids. Several pressing, important questions arise: 
i) are the two chiral spin liquids phases distinct or are they related ? 
ii) is there a simple physical (lattice-based) picture or a variational wave function that describes the chiral spin liquid ? 
iii) what is the "raison d'\^etre" of these chiral spin liquids, i.e. why are the chiral spin liquids stabilized for the two reported Hamiltonians ? Can we come up with some guiding principle which will allow to stabilise CSL on other lattices ? In the following we will address each of these questions. In short we find that the two chiral spin liquids are indeed connected. We then demonstrate that appropriate Gutzwiller projected parton wave functions can have large overlaps with the numerically exact ground states of the studied microscopic models. And finally we show that one location of the chiral spin liquids in parameter space coincides largely with a transition line in the phase diagram of the corresponding classical model. The classical transition line lies between coplanar $\mathbf{q}=0$ magnetic order and a chiral, non-coplanar magnetically ordered phase ({\em cuboc1}~\cite{Messio2012b}).

\begin{figure}[b!]
 \centerline{\includegraphics[width=0.4\linewidth]{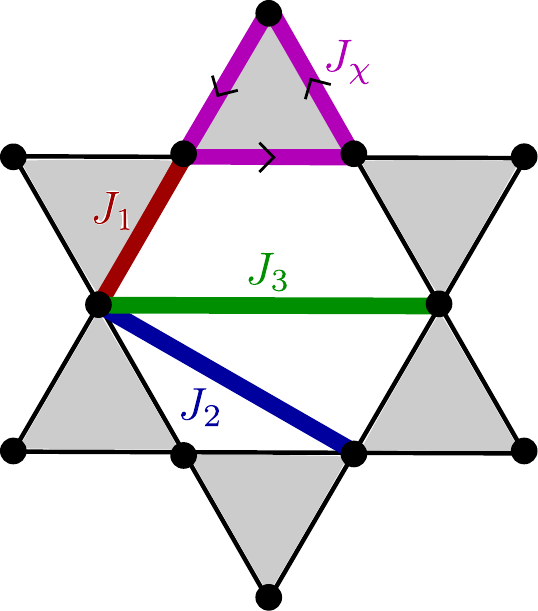}}
  \caption{
   Sketch of the kagome lattice and of the different interaction terms of the Hamiltonian \eqref{eq:chiralhamiltonian}. Heisenberg interactions between first, second and third nearest neighbour are considered. The third nearest neighbour Heisenberg interactions are only considered across the hexagons. Three-spin scalar chirality interactions, breaking time-reversal and parity symmetries, are also considered on grey shaded triangles.
  }
  \label{fig:Hamiltonian}
\end{figure}

\paragraph{Model ---}
We will consider the following Hamiltonian which unifies the two models studied in Refs.~\cite{Gong2014,Bauer2014}:
\begin{equation}
  \label{eq:chiralhamiltonian}
  \begin{aligned}
    H = &J_1 \sum\limits_{\left<i,j\right>} \vec{S}_i \cdot \vec{S}_j + 
    J_2 \sum\limits_{\left<\left<i,j\right>\right>} \vec{S}_i \cdot \vec{S}_j +  \\
    & J_3 \sum\limits_{\left<\left<\left<i,j\right>\right>\right>} \vec{S}_i \cdot \vec{S}_j +
    J_{\chi}\sum\limits_{i,j,k \in \bigtriangleup,\bigtriangledown} \vec{S}_i\cdot(\vec{S}_j\times \vec{S}_k).
  \end{aligned}
\end{equation}

This model includes first, second and third nearest neighbour Heisenberg interactions with coupling constants $J_1$, $J_2$, $J_3$ as sketched in 
Fig.~\ref{fig:Hamiltonian}. The third nearest neighbour Heisenberg interactions are only considered across the hexagons. While these interactions preserve TRS and all the discrete lattice symmetries of the kagome lattice, the additional three-spin scalar chirality interactions on the triangles parametrized by $J_\chi$ break explicitly TRS and spatial parity. Note that Hamiltonian \eqref{eq:chiralhamiltonian} features SU(2) invariance in spin space. For simplicity we will set $J_1 = 1$ in the following.

In Ref.~\cite{Bauer2014} a CSL phase was found for $0.05 \pi \lesssim  \arctan |\frac{J_{\chi}}{J_1}|  \lesssim \pi/2 $ and $J_2=J_3=0$. In this case, TRS is explicitly broken. Interestingly a two-fold degenerate ground state was found, which furthermore exhibits the expected modular data and entanglement spectrum for a topologically ordered chiral $\nu=1/2$ Laughlin state-like phase. On the other hand in Ref.~\cite{Gong2014} a chiral spin liquid with {\em spontaneous} TRS breaking was discovered for $J_{\chi}=0$ and $0.2 \lesssim (J_2 = J_3)/J_1 \lesssim 0.7$. Here the ground state degeneracy is four, which can be understood as arising from two copies of opposite chirality of a two-fold degenerate $\nu=1/2$ Laughlin state. Unlike several topological phases as Toric code \cite{Kitaev20032} and double-semion \cite{Freedman2004428} phases that also have a four-fold ground state degeneracy, we will show that in this case time-reversal symmetry is spontaneously broken.  

\paragraph{Energy spectroscopy ---}

\begin{figure}[t!]
  \centerline{\includegraphics[width=\linewidth]{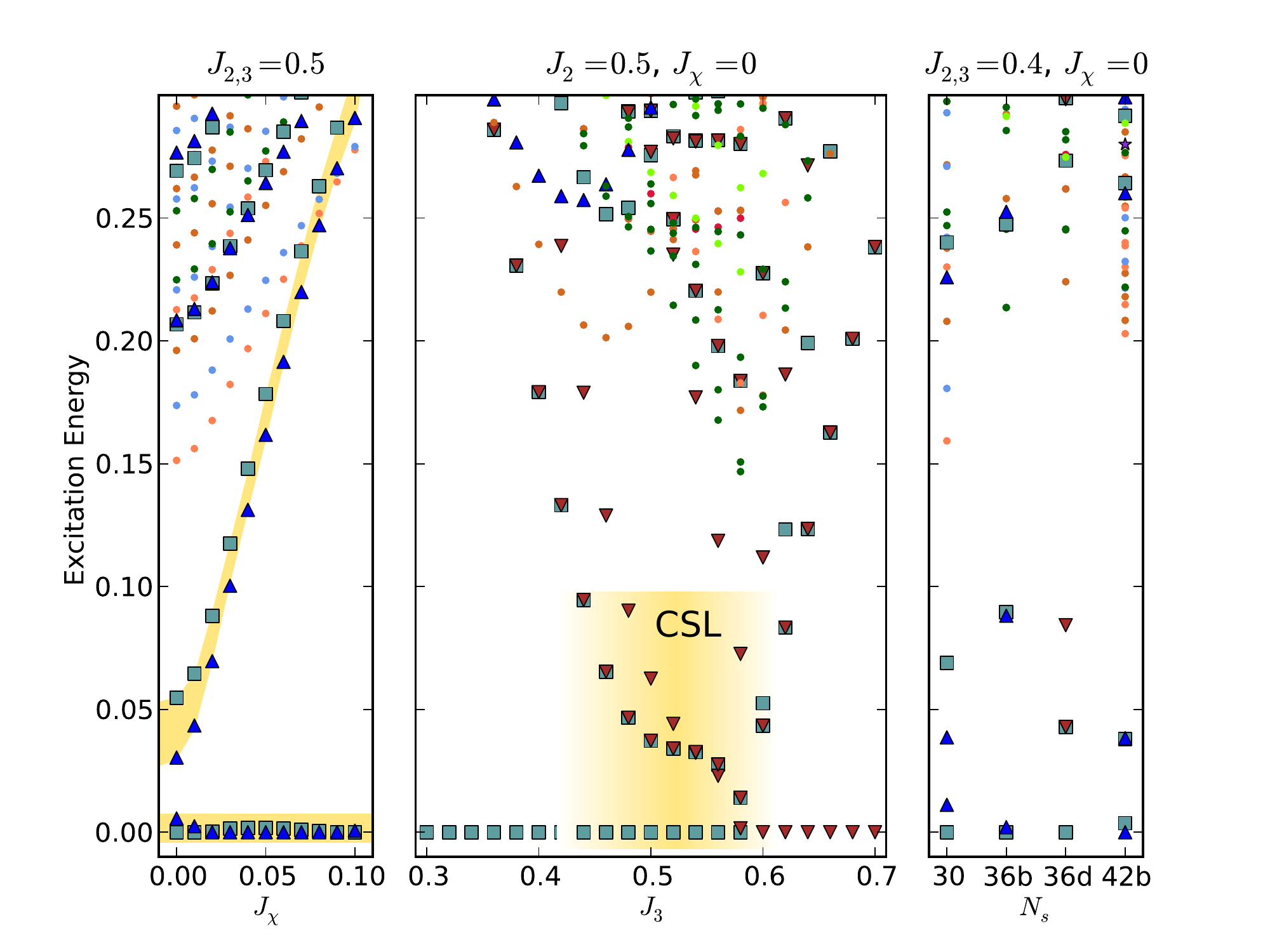}}
  \caption{
  Excitation Spectra from Exact Diagonalization. Different symbols and colors correspond
    to different momentum/pointgroup symmetry sectors. We use the cluster geometries and notation explained in Ref.~\cite{Laeuchli2011} \textbf{(a)} Effect of $J_{\chi}$ term on spectrum on the 30 sites cluster. The four-fold degeneracy of the ground state is lifted to a two-fold degereracy which corresponds to one 
    sign of the scalar chirality. \textbf{(b)} Scan across the classical transition line for $J_{\chi} = 0$ on the 36b sites cluster. The four-fold degereracy of the CSL is only present close to $J_2 = J_3$ (yellow shading). 
    \textbf{(c)} Energy spectra for $J_2=J_3=0.4$, $J_{\chi} = 0$ and various system sizes $N_s$ and geometries.
   \textit{Turquoise rectangle}: 
    $(0,0)\ [\Gamma]$ momentum, even under $180^\circ$ rotation. \textit{Blue up triangle}: $(0,\pi)\ [M]$ momentum, 
    odd under $180^\circ$ rotation. \textit{Red down triangle}: $(0,0)\ [\Gamma]$ momentum, even under $180^\circ$ rotation,
    odd under reflection.}
  \label{fig:edspectra}
\end{figure}

To investigate the persistence of this chiral spin liquid at the thermodynamical limit, we studied the model for $J_2=J_3=0.4$ and $J_{\chi}=0$ up to 42 sites. The low-energy spectra for different system sizes are shown in Fig.~\ref{fig:edspectra} c). While the energy splitting between the four ground states has a non-monotonous behaviour, the energy gap between the four lowest energy states and the fifth one increases with the system size. Moreover, the ratio of the energy splitting to the energy gap  decreases with the system size, this tends to indicate that this phase is indeed realized at the thermodynamical limit. It is also important to notice that the momentum sectors involved in the four-fold degenerate manifold depend on the cluster shape and can be predicted in complete analogy to the Fractional Quantum Hall and Fractional Chern insulator states~\cite{Regnault2011,Laeuchli2013}.

In Fig.~\ref{fig:edspectra} a) we investigate the energy splitting of the four ground states as we switch on a finite $J_\chi$ coupling. At $J_\chi=0$ the long-range order in the spin chirality is spectrally encoded in the presence of two states per topological sector, where the two states have to be at the same momentum, but differ in the spatial reflection quantum number (if the sample allows this symmetry). As is shown in Fig.~\ref{fig:edspectra} a), the two states per sector split very rapidly upon switching on $J_\chi \neq 0$. We can understand the action of $J_\chi$ regarding the scalar chirality in analogy to the effect of a longitudinal magnetic field on the two degenerate ground states in a ferromagnetic Ising model in the ordered phase, where the magnetic field immediately selects one of the two ordered states. As we show later based on overlaps, the chiral spin liquid thus selected by $J_\chi$ is of the same type as the one stabilised in the $J_1-J_\chi$ model alone, and is connected to the TRS symmetric situation in the absence of $J_\chi$.

Finally in Fig.~\ref{fig:edspectra} b) we investigate the effect of a deviation from the $J_2=J_3$ condition (in the absence of $J_\chi$) by fixing $J_2=0.5$ and varying $J_3$. One observes that the four-fold ground state degeneracy is rapidly lifted when $J_3$ deviates more than about $0.05\sim0.1$ from $0.5$. Interestingly the line $0< J_2=J_3 < 1$ is the classical transition line between a magnetically ordered $\mathbf{q}=0$ ground state for $J_3<J_2$ and the non-coplanar magnetically ordered {\em cuboc1} phase for $J_3>J_2$~\cite{Messio2012}. Below we will show that also the overlaps with the variational wave functions are large only in the direct vicinity of this classical transition line. A deeper understanding of the classical ground state configurations on that line and of the effect of quantum fluctuations on that manifold might thus lead to an identification of the crucial ingredients required to predict and uncover chiral spin liquids in TRS Hamiltonians on different lattices.
We note in passing that the explicitly TRS breaking Hamiltonian~\eqref{eq:chiralhamiltonian} with $J_\chi\neq0$ can be considered as a truncated version of a parent Hamiltonian 
for the CSL constructed in Ref.~\cite{Nielsen2012}, similar to the spin Hamiltonian on the square lattice considered in Ref.~\cite{Nielsen2013}.

\paragraph{Parton construction and overlaps ---}
\begin{figure}[t]
  \centerline{\includegraphics[width=0.8\linewidth]{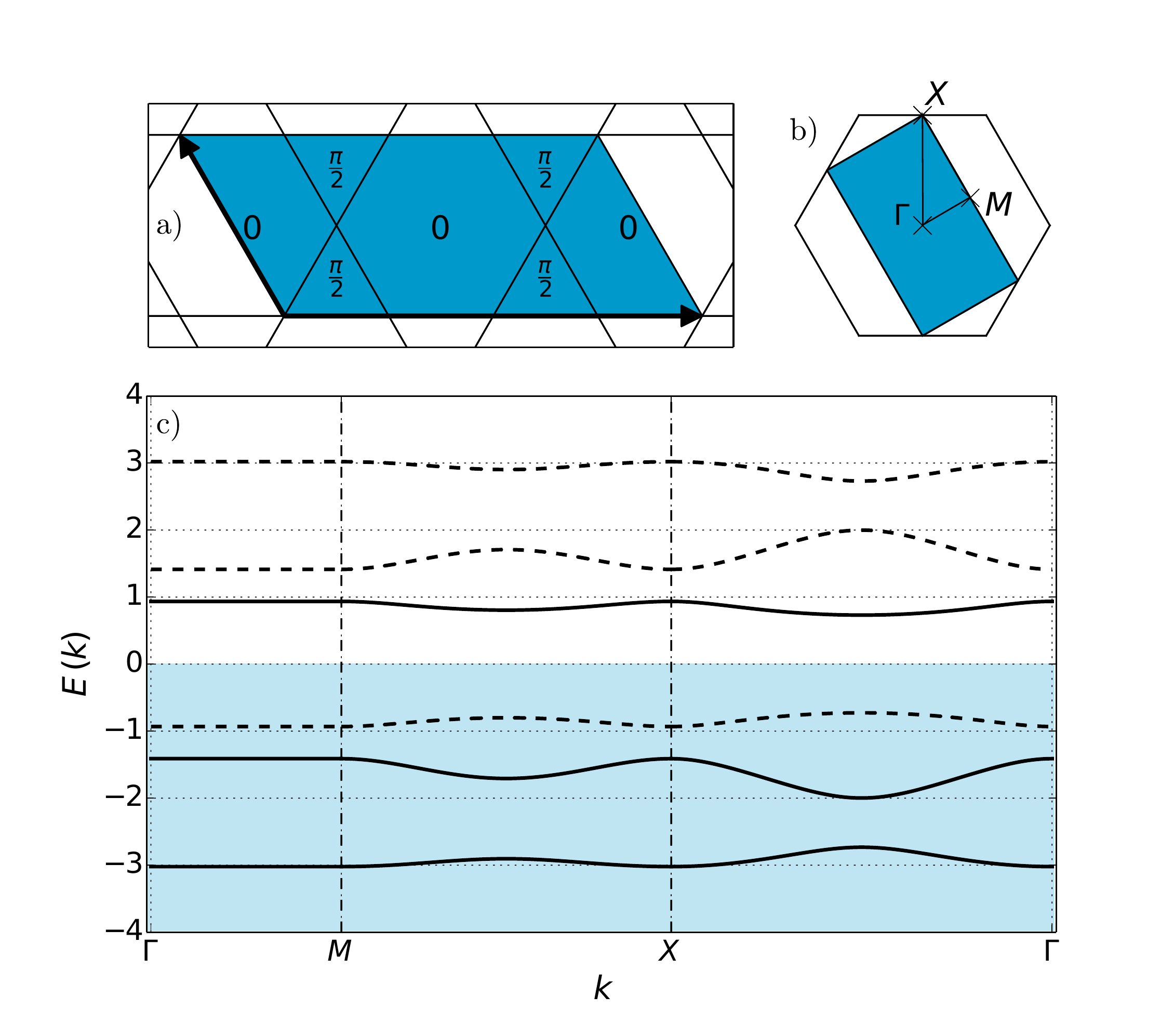}}
  \caption{
  Band structure of the $[\pi/2, 0]$ - model used to generate the pair of model states with positive scalar chirality.
   {\bf a)} geometry of the unit cell. The mean-field parameters $\chi_{ij}$ are chosen such that there are $\pi/2$ fluxes through the triangles and no flux through the hexagons. 
   {\bf b)} Brillouin zone of the model (shaded) and conventional first Brillouin zone of the kagome lattice (hexagon). 
   {\bf c)} Band structure of the model along the path between high symmetry points in the Brillouin zone as drawn in b). 
   The bands are separated by a finite gap. Each solid (dashed) band carries Chern number $-1$ ($+1$).}
  \label{fig:bandstructure}
\end{figure}

As stated earlier on, the CSL can be considered as lattice analogues of the bosonic $\nu=1/2$ Laughlin state. In recent years substantial activity focused on realizing such states on fractionally filled Chern insulators, so called Fractional Chern Insulators (FCI)~\cite{Sheng2011,neupert-PhysRevLett.106.236804,Regnault2011}. It is thus a natural question whether the CSLs under consideration might also have such an interpretation. The natural bosonic $\nu=1/2$ FCI state on the kagome lattice~\cite{2011arXiv1102.2406M,2014arXiv1409.2171K} however does not have the correct magnetisation since it corresponds to magnetisation $m/m_\mathrm{sat}=\pm2/3$ instead of the required $m=0$~\footnote{An interesting idea for future study might be to combine a {\em featureless} Mott insulator wave function with bosonic density $n=1/3$ (i.e. magnetisation $m=2/3$) with a $\nu=1/2$ FCI state.}.

In the absence of a simple FCI candidate wave function we pursue an alternative approach, based on a parton construction.
In order to understand and classify the different spin liquids a generalized construction scheme called parton construction has been introduced by Refs.~\cite{Baskaran1987973,baskaran1988gauge,PhysRevB.38.745,PhysRevB.38.2926,PhysRevLett.76.503,senthil2000z} - see \cite{Wen2004} for an introduction. 
The main idea of this technique is to split up each spin operator $\vec{S}_i$ at site $i$ into two fermionic parton operators $c_{i,\uparrow}$, $c_{i,\downarrow}$ according to 
\begin{equation}
  \begin{aligned}
    S^+_i = c_{i,\uparrow}^\dagger c_{i,\downarrow}\textnormal{, } S^-_i = c_{i,\downarrow}^\dagger c_{i,\uparrow}, \\
    S^z_i = \frac{1}{2}(c_{i,\uparrow}^\dagger c_{i,\uparrow} - c_{i,\downarrow}^\dagger c_{i,\downarrow}) .
  \end{aligned}
\end{equation}
Note that by introducing these operators the Hilbert space is enlarged due to the possibility of doubly occupied or vacant sites. Substituting the parton operators for the spin operators and performing a mean-field approximation by introducing mean-field parameters $\chi_{ij} = \sum_{\sigma}\left < c_{i\sigma}^\dagger c_{j\sigma} \right >$ yields (ignoring constants) a tight-binding model of type 
\begin{equation}
  \label{eq:hbtypepartonmeanfield}
  H_{\textnormal{mean}} = \frac{1}{2} \sum\limits_{i,j,\sigma}  \left(\chi_{ij} c_{i\sigma}^\dagger c_{j\sigma}  +
      \textnormal{h.c.}\right ).  
\end{equation}

Several of these models have been investigated for the kagome lattice~\cite{Marston1991,Hastings2000,Ran2007,Hermele2008,Iqbal2013}. Here we focus on nearest neighbour $\chi_{ij}$ only and the norm is chosen to be $|\chi_{ij}| = 1$. Physically different states can be created by choosing $\chi_{ij}$ such that different magnetic fluxes thread the triangles and the hexagons of the kagome lattice. Amongst these states we consider states whose parent mean-field models have uniform $\pm\pi/2$ flux through the triangles and zero flux through the hexagons~\cite{Marston1991,Hastings2000,Ran2007}. To do so a magnetic six sites unit cell is needed instead of the standard three sites unit cell of the kagome lattice. In the following we will call these the $[\pm\pi/2, 0]$ - models.

The unit cell geometry, Brillouin zone and band structure of the $[\pi/2, 0]$ - model are shown in Fig.~\ref{fig:bandstructure}. All six bands have non-zero Chern numbers as indicated in Fig.~\ref{fig:bandstructure}c). To obtain a $S_z = 0$  model state, the three lowest bands are completely filled both for spins up and spins down and an exact Gutzwiller projection is applied to project
onto the physical spin subspace.  As the filled bands are separated by a finite gap from the empty ones, the spin-spin correlations after projection are expected to decay exponentially with distance, and thus describe a spin disordered state. The Chern number of the filled bands for the $[\pi/2, 0]$ - model ($[-\pi/2, 0]$ - model)  is $-1$ ($+1$). The $[\pi/2, 0]$ - model ($[-\pi/2, 0]$ - model) yields a positive (negative) scalar chirality expectation value for every basic triangle. 
 
On the torus there are two independent non contractible loops. Some of the gauge choices which leave the flux through the triangles and hexagons invariant, correspond to different fluxes through these torus loops. Threading flux through these loops corresponds to a Laughlin flux insertion. Thereby different topological states can be generated. These states cannot be distinguished by local observables and therefore are degenerate for local Hamiltonians in the thermodynamic limit. For the chiral spin liquid a two-fold topological ground state degeneracy is expected. Thus, by threading different fluxes through the torus we should only be able to create a two-dimensional space.
We numerically computed the Gutzwiller projected wavefunctions (GPWFs) of the $[\pm\pi/2, 0]$ - models with a fixed gauge. In order to construct the topological partners of these states we additionally thread fluxes through the torus as explained in the previous section. We checked that for each of the $[\pm\pi/2, 0]$ - models, only two linearly independent states can be only constructed as expected for a CSL within a numerical accuracy of $10^{-3}$, similar as in Ref.~\cite{2014arXiv1409.5427M}.

\begin{figure*}[t!]
  \centerline{\includegraphics[width=\linewidth]{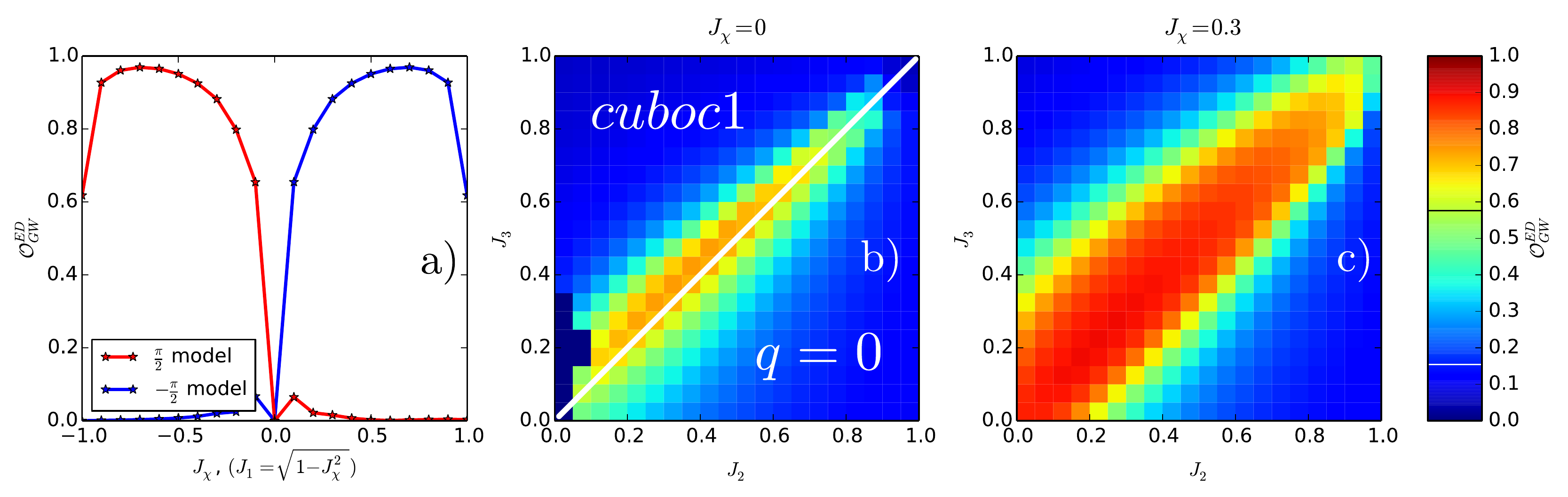}}
  \caption{Overlaps $\mathcal{O}_{\textnormal{GW}}^{\textnormal{ED}}$ of GPWFs with ground states from 
    Exact Diagonalization for $J_2 = J_3 = 0$ and $J_1 = \sqrt{1 - J_{\chi}^2}$ on a 30 sites lattice. 
    The overlaps of the GPWFs of the $\pm\pi/2$-models are symmetric under changing the sign of $J_\chi$. 
    The maximum overlap is equal to $0.97$ and is reached for $J_\chi = \pm 0.7$. The CSL phase extends almost up to the Heisenberg 
    point.
  }
  \label{fig:overlaps_chiral}
\end{figure*}

We compare now these four model states with the ground state $\ket{\psi_{\textnormal{ED}}}$ of the Hamiltonian (\ref{eq:chiralhamiltonian}) obtained using Exact Diagonalization. We choose the overlap $\mathcal{O}_{\textnormal{GW}}^{\textnormal{ED}}$ of the ground state wave function with the four model states as our figure of merit:
\begin{equation}
  \label{eq:overlap}
  \mathcal{O}_{\textnormal{GW}}^{\textnormal{ED}} \equiv \sqrt{\sum_\alpha \left| 
      \braket{\psi_{\textnormal{ED}}}{\psi_{GW}^\alpha} \right| ^2}
\end{equation}
Overlaps of the GPWFs with the ground state of the Hamiltonian (\ref{eq:chiralhamiltonian}) for different parameters on a $N_s=30$ sites sample are shown in Fig.~\ref{fig:overlaps_chiral}. The overlaps of our model state with the ground state wave functions of the model of Ref.~\cite{Bauer2014} where $J_2 = J_3 = 0$, $J_1 = \cos \theta$ and $J_{\chi} =  \sin \theta$ are shown in
Fig.~\ref{fig:overlaps_chiral}(a). We found that overlaps for $J_\chi$ between $0.1$ and $1$ range from $0.62$ to $0.97$. The overlap of the two GPWFs of the $[-\pi/2, 0]$-model are by orders of magnitude larger than those of the $[+\pi/2, 0]$-model. For $J_\chi$ between $-1$ and $-0.1$ the overlaps are exactly the same within numerical precision  as for $J_\chi$ between $0.1$ and $1$ but the role of the GPWFs from the $[+\pi/2, 0]$-model and $[-\pi/2, 0]$-model are exchanged. This is expected since the model with negative $J_\chi$ should have a positive scalar chirality and therefore only little overlap with the variational states form the $[-\pi/2, 0]$-model with negative chirality and vice versa.

For the time-reversal symmetric model with $J_{\chi} = 0$, our variational wave functions have substantial overlap only close to the line $J_2 = J_3$, in agreement with the energy spectroscopy results discussed above [Fig.~\ref{fig:overlaps_chiral}(b)]. In this region the overlaps reach up to $0.72$ for $N_s=30$.

As can be seen in Fig.~\ref{fig:overlaps_chiral} c) for $J_{\chi} = 0.3$ and for $J_{\chi} = 0.6$ (not shown), the region of the CSL broadens significantly when $J_{\chi}$ is increased from zero. For $J_{\chi} = 0.3$ (resp. $J_{\chi} = 0.6$) the overlaps on the classical transition line for $J_2=J_3$ between $0$ and $0.7$ range from $0.8$ to $0.9$ (resp. from $0.85$ to $0.95$).

\paragraph{Conclusion ---}

We showed that the two recently found realizations of chiral spin liquids on the kagome lattice~\cite{Gong2014,Bauer2014} are indeed
related and can be described by Gutzwiller projected parton wave functions. This yields an intuitive microscopic picture of the CSL phase 
stabilized in these models. The ansatz wave functions we chose have been shown to describe a CSL on the kagome lattice~\cite{Marston1991,Hastings2000,Ran2007}. We constructed a pair of Gutzwiller projected parton CSL wave functions for each sign of the scalar chirality. We suggested that these states describe the CSL ground state found on the kagome lattice. To prove that indeed these wavefunctions describe the novel CSL phases found in Refs.~\cite{Gong2014,Bauer2014} we computed overlaps of these variational wave functions with the ground state wave functions computed by Exact Diagonalization. Substantial overlaps were found in regions of the phase diagram where the CSL is expected. By further investigation of excitation spectra, we showed that the CSL phase in Ref.~\cite{Gong2014} is only present on the transition line between a chiral \textit{cuboc1} and a coplanar $q=0$ phase of the classical phase diagram~\cite{Messio2012}. This could serve as a guiding principle for finding CSL phases in other models and on others lattices. Being related to the Laughlin state, these states should exhibit anyonic excitations. Their investigations will be pursue in a future work.

\paragraph{Note added ---} While completing the present manuscript we became aware of parallel work reaching similar conclusions using complementary methods~\cite{Hu2015,Gong2015}.

\acknowledgments
We acknowledge inspiring discussions with H.-H. Tu and A.B. Nielsen.
AW acknowledges support through the FWF project I-1310-N27 (DFG FOR1807).
AS acknowledges support through the FWF SFB Focus (F-4018-N23). This work was supported by the Austrian Ministry of Science BMWF as part of the Konjunkturpaket II of the Focal Point Scientific Computing at the University of Innsbruck.

\bibliography{chiralspinliquid}

\end{document}